\documentclass[aps,prl,twocolumn,showpacs]{revtex4}

\usepackage[dvips]{graphicx}
\usepackage{amssymb}
\usepackage{amsmath}

\begin{document}

\title{Rate effects on layering of a confined linear alkane}

\author{Lionel Bureau}
 \email{bureau@insp.jussieu.fr}
\affiliation{Institut des Nanosciences de Paris, UMR 7588 CNRS-Universit\'e Paris 6, 140 rue de Lourmel, 75015 Paris, France}

\date{\today}

\begin{abstract}

We perform drainage experiments of a linear alkane fluid (n-hexadecane) down to molecular thicknesses, and focus on the role played by the confinement rate. We show that 
molecular layering is strongly influenced by the velocity at which the confining walls are approached:  under high enough shear rates, the confined medium
behaves as a structureless liquid of enhanced viscosity for film thickness below $\sim$10 nm. Our results also lead us to conclude that a rapidly confined film can be quenched in
a metastable disordered state, which might be related with recent intriguing results on the shear properties of 
confined films produced at different rates [Zhu and Granick, Phys. Rev. Lett. {\bf 93}, 096101 (2004)]. 

\end{abstract}


\maketitle


When liquids are confined to molecular thicknesses, their properties deviate markedly from those of the bulk state \cite{Ibook,GBM}. If two solid surfaces are approached
at distances of a few nanometers in a liquid, so-called structural forces are observed, which are associated with the ordering of molecules into layers parallel to the confining walls 
\cite{Ibook,HI,HI2,HKC1,GLL,PRF}. Understanding how such a structuration affects the shear response of confined liquids has been the scope of
 a large number of studies, both numerical \cite{TRG,CMCC,JHT} and experimental. The latter have adressed the issue of flow in nanometer-thick films
using the Surface Force Apparatus (SFA) \cite{CH,CB,HI3,DG,RDG,KK1,Mug} or Atomic Force Microscopy \cite{JPA,Oshea}.

Most of the works mentioned above focus on the lubrication or viscoelastic properties of equilibrated films, {\it i.e.} which have reached their equilibrium layered structure before
shear begins. However, in the course of a recent debate on reassessement of SFA measurements \cite{ZG,ZG2,debate}, 
it has been suggested that the velocity at which a liquid is brought to molecular thickness
may strongly influence its shear response \cite{ZG,ZG2}. In their work, Zhu and Granick \cite{ZG,ZG2} find that thin films formed by rapid confinement 
exhibit a much higher effective viscosity than those produced quasistatically. The authors attribute this effect to the fact that rapid confinement yields less structured films. 

This raises two questions: (i) how does molecular layering depend on the confinement rate, and (ii) if a more disordered film
is produced by rapid confinement, what is the stability of such a ``mechanically quenched'' state ? 

These questions have motivated us to revisit drainage experiments, as performed initially by Chan and Horn \cite{CH}, and to study how the thinning of n-hexadecane confined between mica surfaces is affected
by the velocity at which the surfaces are approached.
We show here that layering is indeed most sensitive to the confinement velocity: at high
enough shear rates, {\it structuring is completely hindered}. For thicknesses in the range 3--10 nm the drainage dynamics is akin to that of a liquid whose effective viscosity increases with the level of
confinement. 
Furthermore, our results lead us to conclude that a non-structured film of hexadecane, obtained by quenching rapidly down to $\sim 2$ nm,  
is metastable and relaxes towards the layered equilibrium configuration via a nucleation/growth process.
    

The experiments were performed using a recently developed surface force apparatus \cite{LB}. As in previous versions of the instrument \cite{TW,IT,IA}, two atomically smooth mica sheets are mounted in a 
crossed-cylinder
 geometry (radius of curvature $R\simeq 1$ cm), and the surfaces are immersed in the liquid under study. Our apparatus allows for independent measurements of the normal force $F$ (by means of a
 capacitive load cell of stiffness 31000 N.m$^{-1}$) and intersurface distance $d$ (using multiple beam interferometry) at a rate on the order of 30 Hz. This enables us to obtain $F(d)$ profiles during drainage of the
 fluid, while the mica surfaces are approached by driving the remote point of the loading spring at a prescribed velocity $V$ in the range 0.05--20 nm.s$^{-1}$.
The mica sheets were prepared according to the following protocole. Muscovite mica plates (JBG-Metafix, France) were cleaved down to a thickness of $\sim$ 10 $\mu$m, cut
into 1 cm$^{2}$ samples by means of surgical scissors, and coated on one side with a 40nm-thick thermally evaporated silver layer. The sheets were fixed, silver side down, onto cylidrical glass lenses, using a
UV setting glue (NOA 81, Norland. UV curing was followed by thermal aging at 50$^{\circ}$C for 12 hours in order to suppress visco-elasticity of the glue layer). The mica sheets were 
placed on the lenses so that their cristallographic axis be aligned. Prior to the experiments, each mica sample was recleaved using adhesive tape \cite{FS}, and mounted in the apparatus
so that the region of closest distance between the lenses be free of steps. The surfaces were brought into contact under an argon atmosphere, and the total mica thickness was
deduced from the position of the fringes of equal chromatic order (FECO) using the multilayer matrix method \cite{Heuberg1,LB}. The surfaces were then separated and the thickness of each mica sheet
determined using the same method. A drop of liquid ($\sim$30 $\mu$L) was finally injected between the surfaces, a beaker containing P$_{2}$O$_{5}$ was placed inside the apparatus which was then sealed
and left for thermal equilibration for about 12 hours before measurements began.
The liquid used was the linear alkane n-hexadecane (Sigma-Aldrich, $>$99\%). The product was filtered through a 0.2 $\mu$m membrane immediately before injection.
All the experiments reported below have been performed at  $T=24 \pm 0.02^{\circ}$C.


We present results obtained in the following conditions. Starting from an initial distance of 20--30 nm, the remote
point of the normal loading spring was driven at a given speed $V$, until a normal force on the order of 500 $\mu$N was reached. Elastic flattening of the surfaces was observed
for loads larger than 100 $\mu$N, and $F=$500  $\mu$N yields a normal pressure of about 500 kPa. 

\begin{figure}[htbp]
$$
\includegraphics[width=8cm]{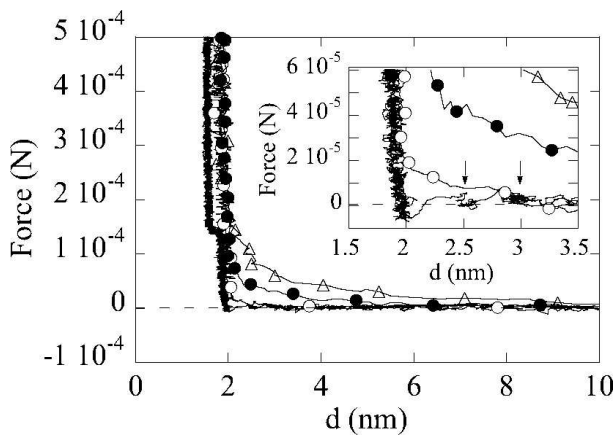}
$$
\caption{Force $vs$ distance curves measured during approach of the mica surfaces at $V=0.05$ nm.s$^{-1}$ (full line); $V=1$ nm.s$^{-1}$ (full line with $\circ$ markers);
$V=5$ nm.s$^{-1}$ (full line with $\bullet$ markers); and $V=10$ nm.s$^{-1}$ (full line with $\triangle$ markers). Inset: close-up view of the same data set. Arrows indicate the first layering transition
detected, between $d\simeq$3 and 2.5 nm.
}
\label{fig:fig1}
\end{figure}

Figure \ref{fig:fig1} shows three $F(d)$ profiles measured at $V=0.05$, 1, 5 and 10 nm.s$^{-1}$. Layering of hexadecane is clearly observed at $V=0.05$ nm.s$^{-1}$: for distances
$d\leq 3$ nm, the film thickness decreases by steps of 4--5 \AA~as the force increases, in agreement with previous studies \cite{HKC1,GLL}. Such a steplike thinning of the 
fluid 
totally disappears for $V\geq 1$ nm.s$^{-1}$ to yield a smooth monotonic repulsive profile (see Fig. \ref{fig:fig1}). 
Besides, we note that for a given film thickness, the normal force is all the larger as the confining velocity is high.

So, for high enough approach velocities, layering is dynamically hindered, and the confined medium stays structureless. 

\begin{figure}[htbp]
$$
\includegraphics[width=8cm]{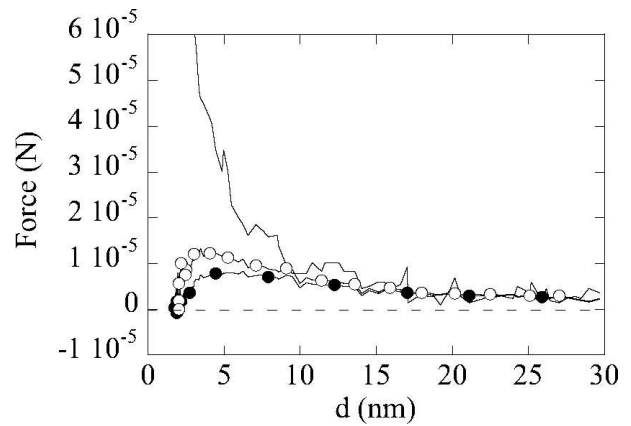}
$$
\caption{Force $vs$ distance curve measured during approach of the mica surfaces at $V=10$ nm.s$^{-1}$ (full line). Hydrodynamic force $F_{H}$ calculated
using expression (\ref{eq:eq1}) and a no-slip boundary condition (full line with $\bullet$). $F_{H}$ calculated using a no-slip plane shifted inside the gap by $d_{s}=0.8$ nm (full line with $\circ$). 
}
\label{fig:fig2}
\end{figure}

Now, at large $V$, hydrodynamic forces may account for a non-negligible part of the measured repulsive forces. 
On fig. \ref{fig:fig2}, we compare the force-distance curve measured at $V=10$ nm.s$^{-1}$ with that corresponding to the squeeze flow of hexadecane, using the bulk 
viscosity $\eta=3.5\times 10^{-3}$ Pa.s. The hydrodynamic force, assuming no-slip boundary conditions, is given by \cite{CH}:
\begin{equation}
F_{H}=6 \pi R^{2} \eta\frac{\dot{d}}{d}
\label{eq:eq1}
\end{equation}
where $R=1$ cm is the radius of curvature of the cylindrical lenses, and $\dot{d}$ is the actual approach velocity of the surfaces. Comparison of $F_{H}(d)$ and $F(d)$ is performed 
for $F<100\, \mu$N, to ensure that
the radius $R$ is not affected by flattening of the surfaces. Fig. \ref{fig:fig2} shows that for $d>$10 nm, the repulsive force can be ascribed to hydrodynamic
flow of the bulk liquid, whereas for $d\lesssim 10$ nm, the measured
forces are larger than $F_{H}$. A modification of eq.(\ref{eq:eq1}), as suggested by Chan and Horn \cite{CH}, to account for shifted no-slip boundary conditions --- 
{\it i.e.} replacing $d$ by $d-2d_{s}$, where $d_{s}$ is the thickness
of an immobile layer near each wall --- does not allow either to fit the experimental $F(d)$ profiles at short distances. 

\begin{figure}[htbp]
$$
\includegraphics[width=8cm]{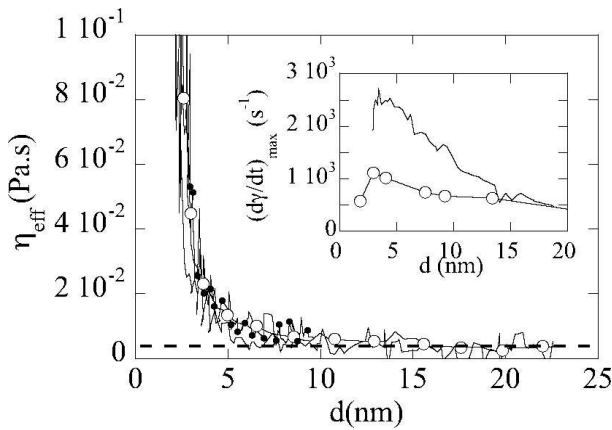}
$$
\caption{Effective viscosity as a function of film thickness for $V=2$ nm.s$^{-1}$ (line with $\bullet$ markers), $V=5$ nm.s$^{-1}$ (full line), and $V=10$ nm.s$^{-1}$ (full line with $\circ$ markers). 
The dashed line indicates the bulk value $\eta=3.5\times 10^{-3}$ Pa.s. Inset: maximum shear rate $\dot{\gamma}_{max}$ $vs$ distance for 
 $V=10$ nm.s$^{-1}$ (full line) in our experiment, and that deduced from Fig.9 of Ref. \cite{CH}, for which $V=16$ nm.s$^{-1}$ (line with $\circ$ markers).}
\label{fig:fig3}
\end{figure}

This points to the existence of a thickness regime in which the non-structured confined fluid does not exhibit bulklike flow properties. 
Tentatively, we use expression (\ref{eq:eq1}) to estimate an effective viscosity
of confined hexadecane from the experimental $F(d)$ data: $\eta_{\text{eff}}=Fd/(6\pi R^{2} \dot{d})$. This is plotted on Fig. \ref{fig:fig3} as a function of thickness, for three driving velocities. 
It is seen that comparable profiles for $\eta_{\text{eff}}(d)$ are obtained, independently of $V$,
which gives support to our analysis in terms of an effective viscosity. Moreover, it appears  that $\eta_{\text{eff}}$ starts to deviate noticeably from the bulk value at short distances, and increases by more 
 than one order of magnitude as $d$ goes roughly from 10 to 3 nm.  
 
This is in apparent contradiction with the results of Chan and Horn \cite{CH}, who found that, assuming the existence of two immobile layers on each wall, bulk viscosity could account for the fluid drainage 
down to thicknesses of 2--3 nm. However, it is important to note here that our apparatus has a stiffness which is more than two orders of magnitude larger than that used in reference \cite{CH}.
From the force balance $F_{H}=K(V t - [d-d_{\text{t=0}}])$, where the right-hand-side term is the restoring force of the cantilever spring of stiffness $K$,  it follows that for a given distance $d$ and 
driving velocity $V$, 
the lower $K$, the lower the approach velocity $\dot{d}$, as well as the maximum shear rate to which the fluid is submitted \cite{CH}: 
 $\dot{\gamma}_{\text{max}}\approx \sqrt{R/d}\times \dot{d}/d$ ($\dot{\gamma}$ reaches its maximum value at a lateral distance of a few microns away from the point of closest approch). 
 In the inset of Fig.\ref{fig:fig3}, we have plotted $\dot{\gamma}_{\text{max}}(d)$ for $V=10$ nm.s$^{-1}$.
 It appears that $\eta_{\text{eff}}$ becomes larger 
 than $\eta_{\text{bulk}}$ when
 $\dot{\gamma}_{\text{max}}\gtrsim 10^3$~s$^{-1}$. An analysis of the data of Chan and Horn in order to evaluate $\dot{\gamma}_{\text{max}}$ shows that, in
 their experiments, $\dot{\gamma}_{\text{max}}\leq10^{3}$ s$^{-1}$, and that its maximum value is
 reached at distances $d\sim 3$ nm (see inset of Fig.\ref{fig:fig3}). Their study and ours are actually compatible, and the present work highlights the advantage of a high stiffness SFA, which gives
 access to a regime which could not be investigated in previous studies. 
 
 Furthermore, from the shear rate value of 10$^{3}$ s$^{-1}$ at which $\eta_{\text{eff}}>\eta_{\text{bulk}}$, we estimate the relaxation time of hexadecane 
 molecules to be $1/\dot{\gamma}_{\text{max}}\simeq$10$^{-3}$ s when $d< 10$ nm, {\it i.e.} many orders of magnitude larger than the Rouse time of bulk C$_{16}$ chains \cite{Baig}.
 This is in qualitative agreement with previous studies on relaxation in confined fluids \cite{DG}.
 
\begin{figure}[htbp]
$$
\includegraphics[width=8cm]{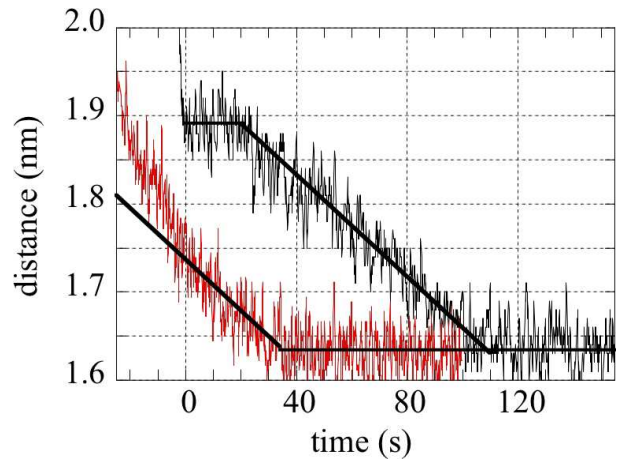}
$$
\caption{Thickness $vs$ time for two films confined at driving velocity $V=10$ nm.s$^{-1}$ (black line), and  $V=1$ nm.s$^{-1}$ (red line). Time $t=0$ corresponds to
the point were loading is stopped and a normal force of 500 $\mu$N is maintained. $t<0$ corresponds to the end of the loading phase. The thick lines are guides for the eye.}
\label{fig:fig4}
\end{figure}

Coming back to Fig. \ref{fig:fig1}, it is clear that under a normal force of 500 $\mu$N, the thickness of the film is 1.65 nm when $V=0.05$ nm.s$^{-1}$ and 1.9 nm
when $V=10$ nm.s$^{-1}$. In order to determine whether the thicker, rapidly quenched, film evolves with time, we have performed the following experiment: 
at $V=10$ nm.s$^{-1}$, hexadecane was confined until a normal force of 500 $\mu$N was reached. The load was then maintained at this value by means of a feedback control loop, and we monitored
 the time evolution of the film thickness $d(t)$, which is plotted on Fig. \ref{fig:fig4}. The $d(t)$ curve exhibits three important features: (i) once the force setpoint has been reached ($t=0$ on 
 Fig. \ref{fig:fig4}), the film thickness stays constant at 1.9 nm for $\sim$20 s; (ii) for $t\gtrsim 20$ s, $d$ decreases {\it linearly} with time ($\dot{d}=3$ pm.s$^{-1}$), until it reaches
 a constant value of 1.65 nm; (iii) the final value of $d$ corresponds to the thickness of 4 molecular layers obtained under quasi-static loading. 
 
 We propose the following picture to interpret such a behaviour. The abrupt stop of the thickness decrease, together with its quasi-linear shape, lead us to rule out relaxation via diffusive-like evacuation of free 
 volume towards the contact edges, which would yield a gradual self-deceleration. On the contrary, such a behavior appears consistent with a nucleation/growth process by which the
 rapidly confined film performs a transition from a higher volume, metastable disordered state, to a lower volume stable layered structure. A similar mechanism has already been evidenced in the
 layer-by-layer collapse of confined liquids \cite{Mug}. In this context, one is tempted to identify the initial 20 s ``incubation'' time with a nucleation delay. However, since our signal is an
 average over a (5 $\mu$m)$^{2}$ zone close to the contact center, we cannot exclude that it might correspond to the time required for the growing layered patch to reach a large enough radius (comparable to the mica thickness
 $\sim 3 \mu$m) for the local bending of the mica sheets to 
 be detectable. 
 
 The above picture is further supported by the following control experiment: hexadecane was confined at a much lower rate, namely $V=1$ nm.s$^{-1}$, and 
 $d(t)$ monitored under $F=$500 $\mu$N. It is seen on Fig. \ref{fig:fig4} that, once the force setpoint is reached, the film thickness decreases from $d=1.75$ nm {\it at the same
 velocity and until the same final value} as those observed in the higher rate experiment. This is consistent with the growth of a stable phase which, under a given applied pressure, occurs at a
 velocity which is independent of the rate at which the film has been produced. Moreover, we note on Fig. \ref{fig:fig4} that no ``nucleation phase'' is observed for $V=1$ nm.s$^{-1}$, and that
 $d$ decreases with time during the loading phase, at a velocity which is about twice that observed under constant load. This suggests that nucleation of the stable phase has occured before reaching the 
 load setpoint, and that propagation, helped by the slowly increasing pressure, takes place at a larger velocity during  loading.

In conclusion, we have performed drainage experiments which unambiguously show that layering of a confined liquid is hindered when the medium is submitted to high enough shear rates. Taking
advantage of a newly developed Surface Force Apparatus, we evidence that the effective viscosity of hexadecane may deviate from its bulk value at thicknesses on the order of 10 nm in situations of
fast out-of-equilibrium confinement.
Moreover, our results indicate that a disordered film formed by a rapid mechanical quench is in a metastable state. Whether the formation of such metastable states is at the origin of
the intriguing results of Zhu and Granick \cite{ZG,ZG2} is still to be clarified.
This calls for further experiments investigating how the yield stress build-up and the steady state shear response of molecularly thin films are affected by the confinement rate.

The author wish to thank C. Caroli and T. Baumberger for fruitful discussions.


\end{document}